\documentclass[%
reprint,
superscriptaddress,
frontmatterverbose,
showpacs,
preprintnumbers,
nofootinbib,
nobibnotes,
amsmath,amssymb,
aps,
prc,
]{revtex4-2}
\usepackage[pdftex]{graphicx}
\usepackage{amsmath}
\usepackage{mathtools}
\usepackage{booktabs}
\usepackage{dcolumn}
\usepackage{bm}
\usepackage[pdftex]{color}
\usepackage{CJK}
\usepackage[T1]{fontenc}
\usepackage{mathrsfs}
\usepackage{multirow}
\usepackage[normalem]{ulem}
\def\ket#1{\left|{#1}\right\rangle}
\def\braket#1#2{\left\langle{#1}\middle|{#2}\right\rangle}
\def\brakket#1#2#3{\left\langle{#1}\middle|{#2}\middle|{#3}\right\rangle}
\def\ve#1{{\bm{#1}}}
\def\nuc#1#2#3{{}^{#2}_{#3}\mathrm{#1}}
\def\urm#1{\scriptstyle{\text{\textrm{\textmd{\textup{#1}}}}}}

\def\avr#1{\left\langle{#1}\right\rangle}

\let\temp\epsilon
\let\epsilon\varepsilon
\let\varepsilon\temp
\let\temp\relax
\let\temp\phi
\let\phi\varphi
\let\varphi\temp
\let\temp\relax

\DeclareMathOperator{\softplus}{softplus}

\begin{document}
% 
%%%%%%%%%%%%%%%%%%%%%%%%%%%%%%%%%%%%%%%%%%%%%%%%%% 
\begin{CJK*}{UTF8}{}
  \preprint{RIKEN-iTHEMS-Report-24}
  \title{A neural network approach for two-body systems with spin and isospin degrees of freedom}
  \author{Chuanxin Wang (\CJKfamily{gbsn}{王传新})}
  \affiliation{
    College of Physics,
    Jilin University,
    Changchun 130012, China}
  \affiliation{
    RIKEN Center for Interdisciplinary Theoretical and Mathematical Sciences Program (iTHEMS),
    Wako 351-0198, Japan}
  \author{Tomoya Naito (\CJKfamily{min}{内藤智也})}
  \email{
    tnaito@ribf.riken.jp}
  \affiliation{
    Department of Nuclear Engineering and Management, Graduate School of Engineering, The University of Tokyo,
    Tokyo 113-8656, Japan}
  \affiliation{
    Department of Physics, Graduate School of Science, The University of Tokyo,
    Tokyo 113-0033, Japan}
  \affiliation{
    RIKEN Center for Interdisciplinary Theoretical and Mathematical Sciences Program (iTHEMS),
    Wako 351-0198, Japan}
  \author{Jian Li (\CJKfamily{gbsn}{李剑})}
  \email{
    jianli@jlu.edu.cn}
  \affiliation{
    College of Physics, Jilin University,
    Changchun 130012, China}
  \author{Haozhao Liang (\CJKfamily{gbsn}{梁豪兆})}
  \email{
    haozhao.liang@phys.s.u-tokyo.ac.jp}
  \affiliation{
    Department of Physics, Graduate School of Science, The University of Tokyo,
    Tokyo 113-0033, Japan}
  \affiliation{
    RIKEN Center for Interdisciplinary Theoretical and Mathematical Sciences Program (iTHEMS),
    Wako 351-0198, Japan}
  \date{\today}
  %%%%%%%%%%%%%%%%%%%%%%%%%%%%%%%%%%%%%%%%%%%%%%%%%% 
  \begin{abstract}
    We propose an enhanced machine learning method to calculate the ground state of two-body systems.
    By extending the original method
    [Naito, Naito, and Hashimoto, Phys.~Rev.~Research \textbf{5}, 033189 (2023)],
    the present method enables consideration of the spin and isospin degrees of freedom by employing a non-fully connected deep neural network and the unsupervised machine learning technique.
    The validity of this method is verified by calculating the unique bound state of the deuteron.
  \end{abstract}
  \maketitle
\end{CJK*}
%%%%%%%%%%%%%%%%%%%%%%%%%%%%%%%%%%%%%%%%%%%%%%%%%% 
% 
\section{Introduction}
\label{sec:level1}
\par
Quantum systems, from atomic nuclei to solids, are composed of many particles, and thus finding states of many-body systems is an essential problem.
In advanced quantum many-body methods, such as density functional theory \cite{
  Hohenberg1964Phys.Rev.136_B864,
  Kohn1965Phys.Rev.140_A1133,
  Kohn1999Rev.Mod.Phys.71_1253},
tensor network \cite{
  White1992Phys.Rev.Lett.69_2863,
  Shinaoka2020SciPostPhys.8_012},
and variational Monte Carlo method \cite{
  McMillan1965Phys.Rev.138_A442,
  Ceperley1977Phys.Rev.B16_3081,
  Ceperley1980Phys.Rev.Lett.45_566},
the ground-state energy can be treated in a variational form, turning finding the ground state of the system into a variational problem.
Among the methods for solving the variational problem, the unsupervised machine learning (ML) is gaining attention due to its powerful performance for optimization, making it an ideal tool to solve many-body problems in quantum systems.
\par
In unsupervised ML for variational problems, the ML structure can be treated as a trial function.
In the case of quantum many-body systems, the energy expectation value $ \avr{H} $ with respect to the system Hamiltonian $ H $ is treated as the loss function and the output of the ML is usually regarded as the wave function. 
\par
For unsupervised ML applications to spin systems, restricted Boltzmann machines (RBMs) were first employed as a wave function ansatz to address both static and time-evolution problems \cite{
  Carleo2017Science355_602}.
Subsequently, improved algorithms were developed to enhance accuracy \cite{
  Nomura2017Phys.Rev.B96_205152}, and deterministic time-evolution schemes were introduced \cite{
  Carleo2018Nat.Commun.9_5322}. Additionally, the Jastrow-Slater wave function ansatz was adapted for unsupervised ML \cite{
  Stokes2020Phys.Rev.B102_205122}.
The excited-state calculation was presented with two different ans\"{a}tze \cite{
  Choo2018Phys.Rev.Lett.121_167204,
  Nomura2020J.Phys.Soc.Jpn.89_054706}.
\par
In the field of nuclear physics, ML has emerged as a crucial tool with extensive and impactful applications \cite{boehnlein2022colloquium,he2023machine}. 
These applications span the prediction of both nuclear ground- and excited-state properties, including nuclear mass
\cite{PhysRevC.111.014325,PhysRevC.106.L021303,Gao2021,Yuan2024,Ming2022},
charge radius \cite{cao2023predictions,tang2024nuclear}, excited states \cite{liu2025prediction,bai2021description}, $\alpha$ decay \cite{rodriguez2019alpha,rodriguez2019bayesian}, $\beta$ decay \cite{niu2019predictions,li2025selection}, charge density \cite{shang2022prediction,shang2024global,source_chinaxiv_202510_00163}, density functional \cite{yang2023kohn}, nuclear level density \cite{du2024inference,Wang_2024}, ground-state magnetic moments \cite{yuan2021magnetic}, 
photoabsorption cross section \cite{parnes2025nuclearresponsesneuralnetworkquantum},
single-$\Lambda$ hypernuclei \cite{zhang2025machinelearningsinglelambdahypernuclei},
and the distribution of ground-state spin in the two-body random ensemble \cite{liu2024neural}.
In statistical mechanics, the transfer matrix for a spin-glass model was calculated by using an unsupervised deep neural network \cite{
  Yoshino2023Phys.Rev.Research5_033068}.
\par
For many-body systems of bosons or fermions depending on continuous coordinates, incorporating the required (anti)symmetrization into the neural network wave function presents a significant challenge.
In Ref.~\cite{
  Keeble2020Phys.Lett.B809_135743},
a single-layer neural network was used to perform the calculation of deuterons in momentum space with a pre-training process.
In Ref.~\cite{sarmiento2024machine}, further analysis was given on the structure of neural networks and numerical uncertainty.
For bosonic systems, Ref.~\cite{
  Saito2018J.Phys.Soc.Jpn.87_074002} introduced a neural network wave function for studying the Calogero-Sutherland model and Efimov bound states, while Ref.~\cite{
  Pescia2022Phys.Rev.Research4_023138} proposed incorporating a pooling layer to enforce symmetrization.
For fermionic systems, such as electronic structure of atoms and molecules and nuclear structure with an \textit{ab initio} Hamiltonian,
the Jastrow-Slater ansatz is often introduced
to consider the fermion antisymmetrization and the interparticle correlation \cite{
  Pfau2020Phys.Rev.Research2_033429,
  Hermann2020Nat.Chem.12_891,
  Adams2021Phys.Rev.Lett.127_022502,
  Gnech2022Few-BodySyst.63_7,
  Yang2022Phys.Lett.B835_137587,
  Yang2023Phys.Rev.C107_034320,
  Fore2023Phys.Rev.Research5_033062}.
The hidden-fermion technique \cite{
  Moreno2022Proc.Natl.Acad.Sci.USA119_e2122059119},
which is an extension of the Jastrow-Slater ansatz,
was proposed and applied with an unsupervised ML.
This technique was immediately applied to nuclear systems to search for the ground-state energy of $ \nuc{O}{16}{} $ \cite{
  Lovato2022Phys.Rev.Research4_043178}, and both energies and other ground-state properties of $A \le 20$ nuclei \cite{PhysRevLett.133.142501}.
Another proposed method to calculate electronic structure was to map the fermionic system into a spin system \cite{
  Choo2020Nat.Commun.11_2368}.
The recent progress is summarized in Ref.~\cite{
  Hermann2023Nat.Rev.Chem.7_692}.
\par
Among these methods, in Ref.~\cite{
  Pescia2022Phys.Rev.Research4_023138},
the bosonic symmetrization is considered using the pooling layer, while it does not work for fermionic systems.
A Jastrow-Slater wave function is applied in many works to solve fermionic systems \cite{
  Pfau2020Phys.Rev.Research2_033429,
  Hermann2020Nat.Chem.12_891,
  Adams2021Phys.Rev.Lett.127_022502,
  Gnech2022Few-BodySyst.63_7,
  Yang2022Phys.Lett.B835_137587,
  Yang2023Phys.Rev.C107_034320,
  Fore2023Phys.Rev.Research5_033062}; however, it does not work for bosonic systems.
\par
In Ref.~\cite{
  Naito2023Phys.Rev.Research5_033189}, another method using an unsupervised deep neural network (DNN) is proposed, which can solve both bosonic systems and fermionic systems in coordinate space.
The calculation steps are the same for both bosonic and fermionic systems, differing only in a sign in the trial wave function to account for (anti)symmetrization.
According to the universal approximation theorem \cite{
  Hornik1989NeuralNetw.2_359,
  Hornik1991NeuralNetw.4_251},
a multilayer neural network can approximate any functions with desired accuracy with changing the size of the neural network.
Therefore, a neural network can be used as a trial wave function \cite{
  Carleo2017Science355_602,
  Nomura2017Phys.Rev.B96_205152,
  Carleo2018Nat.Commun.9_5322,
  Keeble2020Phys.Lett.B809_135743,
  sarmiento2024machine,
  Saito2018J.Phys.Soc.Jpn.87_074002,
  Hermann2023Nat.Rev.Chem.7_692,
  Naito2023Phys.Rev.Research5_033189}.
By utilizing energy expectation as the loss function for minimization, this method successfully calculated the ground state of one-dimensional (1D) one- to three-body systems.
However, the spin and isospin degrees of freedom, which are crucial for discussing the properties of magnetic materials and atomic nuclei, have not been considered yet in this method.~\footnote{
  It should be noted that the spin and isospin degrees of freedom have been taken into account
  for ML calculation with the Jastrow-Slater ansatz
  in, e.g., Refs.~\cite{
    Lovato2022Phys.Rev.Research4_043178,
    Adams2021Phys.Rev.Lett.127_022502,
    Gnech2022Few-BodySyst.63_7,
    Lovato2022Phys.Rev.Research4_043178,
    Yang2022Phys.Lett.B835_137587,
    Yang2023Phys.Rev.C107_034320,
    Fore2023Phys.Rev.Research5_033062}.}
\par
In this study, we extend the method proposed in Ref.~\cite{
  Naito2023Phys.Rev.Research5_033189}
to include the spin and isospin degrees of freedom.
We will focus on the two-body systems without the external potential as the first step. 
Owing to translational symmetry, we only consider the relative motion. 
This setup corresponds to the one-body calculation in Ref.~\cite{
  Naito2023Phys.Rev.Research5_033189}.
This method is validated through the calculation of the deuteron, the simplest realistic many-body system. 
Its Hamiltonian contains a non-central tensor force, making the inclusion of spin and isospin degrees of freedom essential.
Since a deuteron has only one bound state, we do not focus on the excited states in the present discussions.
\par
Note that the system we consider is the same as in Refs.~\cite{
  Keeble2020Phys.Lett.B809_135743,
  sarmiento2024machine}. 
However, our method is technically different from them.
In the process of optimization, Ref.~\cite{
  Keeble2020Phys.Lett.B809_135743} used a supervised method to pre-train DNN parameters, facilitating faster convergence and ensuring accurate behavior at the origin \cite{sarmiento2024machine, keeble2022neural}.
Our method does not require pre-training, primarily due to the choice of DNN ansatze for the wave functions.
This choice influences the behavior at the origin (discussed in the Appendix), along with other technical differences like mesh point distributions and optimizers.
It was pointed out in Ref.~\cite{sarmiento2024machine} that the structure of a neural network affects the convergence, where deeper neural network gives lower convergence rate.
On the contrary, our DNN model always converges with different numbers of layers by ensuring a relatively much smaller number of hidden nodes.
For results, overfitting is not observed in our DNN wave functions, whereas it appears in Ref.~\cite{sarmiento2024machine}.
In Ref.~\cite{
  Keeble2020Phys.Lett.B809_135743}, it was
also pointed out that the extension to arbitrary spin and isospin was a future perspective;
our work raises the general formula for arbitrary spin and isospin and calculated the deuteron with 18 possible partial waves by applying a non-fully connected DNN structure.
\par
The remaining of this paper is organized as follows: 
In Sec.~\ref{Sect:II}, we discuss the theory to include the spin and isospin degrees of freedom with partial wave expansions.
The corresponding DNN model is proposed in Sec.~\ref{Sect:III}.
In Sec.~\ref{Sect:IV}, we discuss calculation results of deuteron.
Finally, we give the summary and future perspectives in Sec.~\ref{Sect:V}.

%%%%%%%%%%%%%%%%%%%%%%%%%%%%%%%%%%%%%%%%%%%%%%%%%% 
% 
\section{Method}
\label{Sect:II}
%%%%%%%%%%%%%%%%%%%%%%%%%%%%%%%%%%%%%%%%%%%%%%%%% 
% 
\subsection{Hamiltonian for two-body systems with spin and isospin degrees of freedom}
\par
A general self-bound two-body system is studied with the Hamiltonian expressed in coordinate space as
\begin{equation}
  H
  =
  -
  \frac{\hbar^2}{2 m_1}
  \nabla_1^2
  -
  \frac{\hbar^2}{2 m_2}
  \nabla_2^2
  +
  V^{\urm{int}} \left( \ve{r}_1, \ve{r}_2 \right),
\end{equation}
where the $ \nabla^2 $ denotes the Laplacian and the $ V^{\urm{int}} $ denotes the two-body interaction.
Since we consider a self-bound system, which has a translational symmetry,
the center-of-mass motion can be isolated;
hence, only the relative motion is accounted for the following calculation.
Then, the problem is truncated into the one-body system.
By defining the center of mass $ M $ and the reduced mass $ \mu $ as
\begin{equation}
  M
  =
  m_1
  +
  m_2,
  \qquad
  \mu
  =
  \frac{m_1 m_2}{m_1 + m_2}
\end{equation}
with the center-of-mass and relative coordinates
\begin{equation}
  \ve{R}
  =
  \frac{m_1 \ve{r}_1 + m_2 \ve{r}_2}{m_1 + m_2},
  \qquad
  \ve{r}
  = 
  \ve{r}_1
  -
  \ve{r}_2,
\end{equation}
respectively, the Hamiltonian can be split into the center-of-mass part $ H_R $ and relative motion part $ H_r $ as 
\begin{subequations}
  \begin{align}
    H
    & =
      H_R
      +
      H_r,\\
    H_R
    & =
      -
      \frac{\hbar^2}{2 M}
      \nabla_R^2,\\
    H_r
    & =
      -
      \frac{\hbar^2}{2 \mu}
      \nabla_r^2
      +
      V^{\urm{int}} \left( \ve{r} \right).
  \end{align}
\end{subequations}
The center-of-mass part $ H_R $ can be omitted since it describes the behavior of a free particle. 
Thus, only the relative motion part $ H_r $ is considered.
\par
For example, we assume that the interaction between particles depends only on the relative distance and is spherically symmetric.
Then, the wave function can be expanded using spherical harmonics.
The relative motion part of the Hamiltonian of the radial wave function reads
\begin{equation}
  \label{eq5}
  H_r
  =
  -
  \frac{\hbar^2}{2 \mu}
  \left(
    \frac{\partial^2}{\partial r^2}
    +
    \frac{2}{r}
    \frac{\partial}{\partial r}
  \right)
  + \frac{\hbar^2 l \left( l+1 \right)}{2 \mu r^2}
  +
  V^{\urm{int}}\left( r \right)
\end{equation} 
with $ l $ the orbital quantum number.
The ground state energy of $ H_r $ is denoted as $ E_0 $.
% 
%%%%%%%%%%%%%%%%%%%%%%%%%%%%%%%%%%%%%%%%%%%%%%%%%% 
% 
\subsection{Argonne series nucleon-nucleon potentials}
\par
Compared to the interaction between two electrons, nuclear interactions are much more complex, including the isospin dependence and the non-central tensor force.
In the following, we consider a Hamiltonian with the Argonne V18 (AV18) potential \cite{
  Wiringa1995Phys.Rev.C51_38}
to illustrate the present method.~\footnote{
  It should be noted that any local potentials,
  including the state-of-the-arts chiral effective potentials~\cite{
    Gezerlis2013Phys.Rev.Lett.111_032501,
    Roggero2014Phys.Rev.Lett.112_221103,
    Lynn2014Phys.Rev.Lett.113_192501,
    Piarulli2015Phys.Rev.C91_024003,
    Piarulli2016Phys.Rev.C94_054007,
    Lonardoni2018Phys.Rev.C97_044318,
    Lonardoni2020Phys.Rev.Research2_022033},
  can be used in our method,
  while this method can also be applied for the momentum-space representation straightforwardly.}
The AV18 potential is a nucleon-nucleon potential with explicit charge dependence and charge asymmetry,
and it is composed of 18 operators
\begin{equation}
  \label{eq:AV18}
  V^{\urm{int}} \left( r \right)
  =
  \sum_{p = 1}^{18}
  V_p \left( r \right)
  O^p .
\end{equation}
The term $ O^p $ denotes a series of 18 operators.
The first 14 operators are charge independent,
\begin{align}
  & O^{\urm{$ p = 1 $--$ 14 $}}
    \notag \\
  & =
    \left[
    1, \, 
    \left( \ve{\sigma}_1 \cdot \ve{\sigma}_2 \right), \, 
    S_{12}, \, 
    \left( \ve{L} \cdot \ve{S} \right), \, 
    \ve{L}^2, \, 
    \ve{L}^2 \left( \ve{\sigma}_1 \cdot \ve{\sigma}_2 \right), \, 
    \left( \ve{L} \cdot \ve{S} \right)^2
    \right]
    \notag \\
  & \quad
    \otimes
    \left[
    1, \,
    \left( \ve{\tau}_1 \cdot \ve{\tau}_2 \right)
    \right],
\end{align}
and the last four break the charge independence,
\begin{equation}
  O^{\urm{$ p = 15 $--$ 18 $}}
  =
  \left[
    1, \,
    \left( \ve{\sigma}_1 \cdot \ve{\sigma}_2 \right), \,
    S_{12}
  \right]
  \otimes
  T_{12},
  \,
  (\tau_{z1} + \tau_{z2}).
\end{equation}
The operators include the Pauli matrices of spin $ \ve{\sigma} $ and isospin $ \ve{\tau} $,
the $ z $-projection of isospin $ \tau_z $,
the product of orbital angular momentum $ \ve{L} $ and spin $ \ve{S} $,
and
the tensor operator $ S_{12} $ and the isotensor operator $ T_{12} $ defined by
\begin{subequations}
  \begin{align}
    S_{12}
    & =
      3 \frac{\left( \ve{\sigma}_1 \cdot \ve{r} \right) \left( \ve{\sigma}_2 \cdot \ve{r} \right)}{r^2}
      -
      \ve{\sigma}_1 \cdot \ve{\sigma}_2, \\
    T_{12}
    & =
      3 \tau_{z1} \tau_{z2}
      -
      \ve{\tau}_1 \cdot \ve{\tau}_2,
  \end{align}
\end{subequations}
respectively.
The $ V_p \left( r \right) $ are obtained by fitting into 
the two-nucleon scattering data and the deuteron binding energy \cite{
  Wiringa1995Phys.Rev.C51_38}.
On top of the nuclear interaction defined by Eq.~\eqref{eq:AV18} and the proton-proton Coulomb interaction,
the higher-order corrections of the electromagnetic interactions are included;
hence, the electromagnetic interaction even exists between a proton and a neutron.
\par
Besides the AV18 potential,
for the calculation of the deuteron, we will also use the Argonne V8\ensuremath{'} (AV8\ensuremath{'}) \cite{
  Pudliner1997Phys.Rev.C56_1720}
and the Argonne V4\ensuremath{'} (AV4\ensuremath{'}) \cite{
  Wiringa2002Phys.Rev.Lett.89_182501}
potentials.
The AV8\ensuremath{'} potential is a simplified version of the AV18 potential with the number of operators reducing from 18 to 8, 
\begin{equation}
  O^{\urm{$ p = 1 $--$ 8 $}}
  =
  \left[
    1, \, 
    \left( \ve{\sigma}_1 \cdot \ve{\sigma}_2 \right), \, 
    S_{12}, \, 
    \left( \ve{L} \cdot \ve{S} \right)
  \right]
  \otimes
  \left[
    1, \,
    \left( \ve{\tau}_1 \cdot \ve{\tau}_2 \right)
  \right].
\end{equation}
The AV4\ensuremath{'} potential is a further simplified version with the number of operators reducing from 8 to 4,
\begin{equation}
  O^{\urm{$ p = 1 $--$ 4 $}}
  =
  \left[
    1, \, 
    \left( \ve{\sigma}_1 \cdot \ve{\sigma}_2 \right)
  \right]
  \otimes
  \left[
    1, \,
    \left( \ve{\tau}_1 \cdot \ve{\tau}_2 \right)
  \right].
\end{equation}
The radial potentials $ V_p \left( r \right) $ for both AV8\ensuremath{'} and AV4\ensuremath{'} potentials are a linear combination of $ V_p \left( r \right) $ of the AV18 one.
% 
%%%%%%%%%%%%%%%%%%%%%%%%%%%%%%%%%%%%%%%%%%%%%%%%%% 
\subsection{Partial wave expansion of wave function}
\par
To find the ground state of the selected Hamiltonian, we rewrite the quantum state using the partial wave expansion
\begin{align}
  & \ket{\Psi}
    \notag \\
  & =
    \sum_{\substack{L, m_L, S, \\ m_S, T, m_T}}
  a_{L m_L S m_S T m_T}
  \phi_{L m_L S m_S T m_T}
  Y_{L m_L}
  \ket{S m_S}
  \otimes
  \ket{T m_T}.
\end{align}
The $ \phi_{L m_L S m_S T m_T} $ denotes the radial part of the wave function.
The $ Y_{L m_L} $, $ \ket{S m_S} $, and $ \ket{T m_T} $
are the eigenstates of operators
$ \ve{L}^2 $, $ \ve{S}^2 $, and $ \ve{T}^2 $, respectively.
Because the deuteron is an isospin singlet state $\left(T  = 0\right)$,
we only consider $ \ket{T m_T} = \ket{0 0} $ and omit it in the following.
We assume partial wave functions are normalized, i.e., 
\begin{equation}
  \sum_{\substack{L, m_L,\\ S, m_S}}
  a_{L m_L S m_S}^2
  =
  1.
\end{equation}
\par
Because there exists the $ \left( \ve{L} \cdot \ve{S} \right) $ term in the AV18 and AV8\ensuremath{'} potentials,
the quantum numbers $ L \left( L + 1 \right) $ and $ S \left( S + 1 \right) $ are no longer good quantum numbers. 
Therefore, we introduce $ J \left( J + 1 \right) $ as a good quantum number, which is the eigenvalue of the squared total angular momentum operator $\ve{J}^2$, where $ \ve{J} = \ve{L} + \ve{S} $.
By changing the basis of partial waves, the new expansion of the wave function reads
\begin{align}
  \ket{\Psi}
  & =
    \sum_{\substack{L, m_L,\\ S, m_S}}
  a_{L m_L S m_S}
  \phi_{L m_L S m_S}
  Y_{L m_L}
  \ket{S m_S}
  \notag \\
  & =
    \sum_{\substack{L, S,\\ J, m_J}}
  b_{L S J m_J}
  \psi_{L S J m_J}
  \ket{L S J m_J}
\end{align}
with the normalization condition for the new coefficients
\begin{equation}
  \sum_{\substack{L, S,\\ J, m_J}}
  b_{L S J m_J}^2
  =
  1.
\end{equation}
By utilizing the Clebsch-Gordan coefficients
$ c_{j_1 m_{j_1} j_2 m_{j_2}}^{J m_J}$,
we have the expansion
\begin{equation}
  \ket{L S J m_J}
  =
  \sum_{m_L, m_S}
  c_{L m_L S m_S}^{J m_J}
  Y_{L m_L}
  \ket{S m_S}.
\end{equation}
Therefore, the wave function can be simplified as
\begin{equation}
  \ket{\Psi}
  =
  \sum_{\substack{L, S, \\ J, m_J}}
  b_{L S J m_J}
  \psi_{L S J m_J}
  \sum_{m_L, m_S}
  c_{L m_L S m_S}^{J m_J}
  Y_{L m_L}
  \ket{S m_S}.
\end{equation}
Because of the $ SO \left( 3 \right) $ symmetry for $ J $,
the states with different $ m_J $ degenerate
\begin{equation}
  d_{L S J}
  \varphi_{L S J}
  = 
  \color{black}b_{L S J m_J}
  \psi_{L S J m_J}.
\end{equation}
The final partial wave expansion is
\begin{equation}
  \label{eq18}
  \ket{\Psi}
  =
  \sum_{L, S, J}
  d_{L S J}
  \varphi_{L S J}
  \sum_{m_L, m_S, m_J}
  c_{L m_L S m_S}^{J m_J}
  Y_{L, m_L}
  \ket{S m_S}
\end{equation}
with the normalization condition
\begin{equation}
  \sum_{L, S, J}
  d_{L S J}^2
  \left( 2J + 1 \right)
  =
  1.
\end{equation}
% 
%%%%%%%%%%%%%%%%%%%%%%%%%%%%%%%%%%%%%%%%%%%%%%%%%% 
\section{Neural Network Models}
\label{Sect:III}
%%%%%%%%%%%%%%%%%%%%%%%%%%%%%%%%%%%%%%%%%%%%%%%%%% 
% 
\subsection{The discrete form of Hamiltonian and wave functions in neural network}
\par
In the present neural network approach, the wave function is represented by a DNN. 
Because we focus on the bound states of systems only, it is enough to calculate the system within a box of finite size.
The spatial coordinate $ r $ is discretized as the input variables
and the partial wave functions $ \varphi_{L S J} $ as the output variables of DNN,
so that they can be represented as vectors. 
The function $\xi_{LSJ} \left(r \right)= r \varphi_{LSJ} \left(r \right)$ is defined to impose the Dirichlet boundary condition $\xi_{LSJ} \left(0 \right) = \xi_{LSJ} \left(\infty \right) = 0$ for the loss function calculation.
The function $\xi_{LSJ}$ satisfies
\begin{equation}
  H'_r \xi_{LSJ} \left(r \right) = E_0 \xi_{LSJ} \left(r \right),
\end{equation} 
where $ H'_r $ reads
\begin{equation}
  H'_r
  =
  -
  \frac{\hbar^2}{2 \mu}
  \frac{\partial^2}{\partial r^2}
  + \frac{\hbar^2 l \left( l+1 \right)}{2 \mu r^2}
  +
  V^{\urm{int}}\left( r \right).
\end{equation} 
\par 
To perform the calculation numerically, the mesh points of spatial coordinates are evenly distributed with size $ \Delta r $, with $ \left(M + 1\right) $ mesh points in total.
Because the boundary condition is assumed as $\xi\left( 0\right) = \xi\left( r_\text{Max}\right) = 0$, only $ \left(M - 1\right) $ mesh points are calculated.
In detail, the output of the DNN is used as $\varphi_{L S J}$ and the loss function is calculated by using $\xi_{LSJ}$,
since using $\varphi_{L S J}$ as the output gives better accuracy than using $\xi_{LSJ}$ as shown in the Appendix.
The $\varphi\left(0\right)$ is calculated by the Lagrange interpolating polynomial with five adjacent points.
The vector corresponding to the radial partial wave function $ \varphi_{L S J} $ of Eq.~\eqref{eq18} reads
\begin{equation}
  \varphi_{L S J}
  \simeq
  \begin{pmatrix}
    \tilde{\varphi}_{L S J 1} \\
    \tilde{\varphi}_{L S J 2} \\
    \tilde{\varphi}_{L S J 3} \\
    \vdots \\
    \tilde{\varphi}_{L S J \left( M - 3 \right)} \\
    \tilde{\varphi}_{L S J \left( M - 2 \right)} \\
    \tilde{\varphi}_{L S J \left( M - 1 \right)} 
  \end{pmatrix},
\end{equation}
where the $ i $-th component is 
\begin{equation}
  \label{eq22}
  \tilde{\varphi}_{L S J i}
  =
  \varphi_{L S J} \left( r_i \right),
\end{equation}
with $ r_i = i \, \Delta r $.
\par
The second derivative in the $ H'_r $ reads
\begin{equation}
  \frac{\partial^2}{\partial r^2}
  \simeq
  \frac{1}{\left( \Delta r \right)^2}
  \begin{pmatrix}
    -2     &  1     &  0     & \ldots &  0     &  0     &  0     \\
    1      & -2     &  1     & \ldots &  0     &  0     &  0     \\
    0      &  1     & -2     & \ldots &  0     &  0     &  0     \\
    \vdots & \vdots & \vdots & \ddots & \vdots & \vdots & \vdots \\
    0      &  0     &  0     & \ldots & -2     &  1     &  0     \\
    0      &  0     &  0     & \ldots &  1     & -2     &  1     \\
    0      &  0     &  0     & \ldots &  0     &  1     & -2     
  \end{pmatrix}.
\end{equation}
The rest part of $ H'_r $,
denoted as $ V $ altogether,
are expressed as diagonal matrices
\begin{equation}
  V
  \simeq 
  \begin{pmatrix}
    \tilde{V}_1 & 0           & 0           & \ldots & 0               & 0               & 0               \\
    0           & \tilde{V}_2 & 0           & \ldots & 0               & 0               & 0               \\
    0           & 0           & \tilde{V}_3 & \ldots & 0               & 0               & 0               \\
    \vdots      & \vdots      & \vdots      & \ddots & \vdots          & \vdots          & \vdots          \\
    0           & 0           & 0           & \ldots & \tilde{V}_{M-3} & 0               & 0               \\
    0           & 0           & 0           & \ldots & 0               & \tilde{V}_{M-2} & 0               \\
    0           & 0           & 0           & \ldots & 0               & 0               & \tilde{V}_{M-1} 
  \end{pmatrix},
\end{equation}
where $ \tilde{V}_i = V \left( r_i \right) $.
The ground-state energy $ E_0 $ can then be calculated by
\begin{equation}
  \label{eq26}
  E_0
  \simeq
  \frac{\brakket{\Psi}{H}{\Psi}}{\braket{\Psi}{\Psi}}.
\end{equation} 
% 
%%%%%%%%%%%%%%%%%%%%%%%%%%%%%%%%%%%%%%%%%%%%%%%%%% 
\subsection{The design of the non-fully-connected neural network}
\par
\begin{figure}[tb]
  \includegraphics[width=1.0\linewidth]{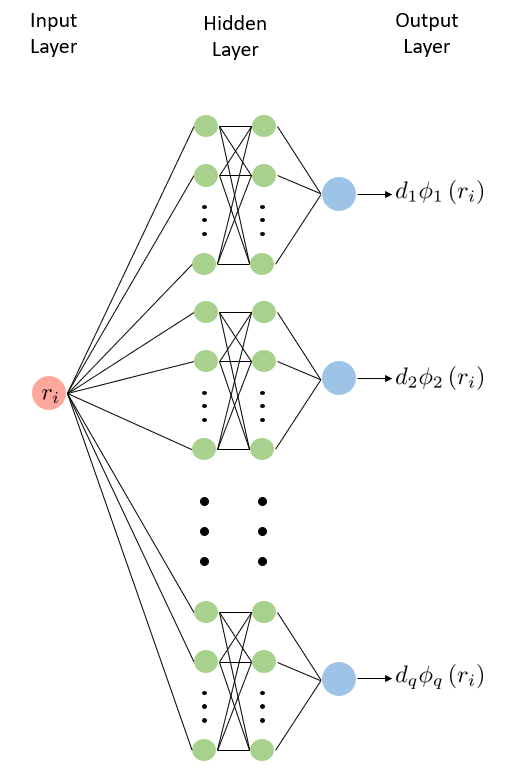}
  \caption{
    Schematic figure of the non-fully-connected deep neural network.
    The shared input is the relative distance between the particles.
    The outputs are the partial wave functions with coefficients $d_j \varphi_j $,
    where $ j $ ($ j = 1 $, $ 2 $, \ldots, $ q $)
    represents the combination of the angular momenta, $ L $, $ S $, and $ J $.
    There is no connection between hidden layer nodes of different outputs.
    The computational accuracy can be enhanced by increasing the number of layers and nodes in each layer.}
  \label{fig:schematic}
\end{figure}

The design of the present non-fully-connected DNN, which generates partial wave functions with corresponding coefficients $d_j \varphi_j $,
is shown in Fig.~\ref{fig:schematic}. 
The input is the relative distance $ r_i $.
The outputs are the partial wave functions.
After determining the cutoff value of the orbital angular momentum quantum number $ L $ and the spin quantum number $ S $,
the possible angular momentum sets $ \left\{ L S J \right\} $ of partial waves are then determined by the $ LS $ coupling.
For each single output, the nodes in the hidden layers are fully connected,
while the hidden layer nodes of different outputs are not connected to each other.
The ``softplus'' function 
\begin{equation}
  \softplus
  \left( x \right)
  =
  \log
  \left(
    1
    +
    e^x
  \right)
\end{equation}
is used as the activation function.
\par
Throughout the training process, the relative distances of the particles are discretized into $ M $ uniformly distributed lattice points, which are passed to the DNN as a dataset.
The output of the DNN is a dataset containing all the partial wave functions.
The energy expectation value is regarded as the loss function, which is calculated from Eq.~\eqref{eq26}.
The parameters of DNN are then updated by the Adam optimizer~\cite{
  DBLP:journals/corr/KingmaB14}
and the ML is implemented by using the \textsc{Tensorflow}~\cite{
  tensorflow2015-whitepaper}.
% 
%%%%%%%%%%%%%%%%%%%%%%%%%%%%%%%%%%%%%%%%%%%%%%%%%% 
% 
\section{Calculation results for deuteron}
\label{Sect:IV}
%%%%%%%%%%%%%%%%%%%%%%%%%%%%%%%%%%%%%%%%%%%%%%%%%% 
% 
\par
The feasibility of the method is validated by applying it to deuteron.
To include a more general case, at the beginning, we consider the orbital angular momentum $ L $ from zero to four
and spin $ S $ from zero to one in Eq.~\eqref{eq22}, with 18 possible states in total.
The 18 partial wave functions are computed in a $ 20 \, \mathrm{fm} $ box with $ 1500 $ mesh points.
A non fully-connected DNN is used, which is composed of two layers with each containing $ 16 $ units.
Moreover, we also test a fully-connected version of such DNN to explore the impact of connections between nodes on the training results.
Both models are trained to testify whether only the $ {}^3 S_1 $ and $ {}^3 D_1 $ states are obtained,
which is a known fact for the properties of deuteron \cite{Bertulani2007NuclearPhysicsinaNutshell_PrincetonUniversityPress}.
Table~\ref{tab:table1} lists the percentage of each state in the wave function.
The results show that only the $ {}^3 S_1 $ and $ {}^3 D_1 $ states make significant contributions in the ground state,
while the contributions of the other states are small enough to be neglected, which is consistent with the known fact.
It is also found that both the fully-connected and non-fully-connected DNN can generate the ground state on the same level of accuracy, showing that the connection between nodes does not make significant difference on the precision, and we choose the non-fully connected one for the following calculations.
The relative error of the ground-state energy is $ 0.05 \, \% $ of the benchmark value
obtained by the Green's function Monte Carlo calculation~\cite{
  Wiringa1995Phys.Rev.C51_38}. 
\par
\begin{figure}[tb]
  \includegraphics[width=1.0\linewidth]{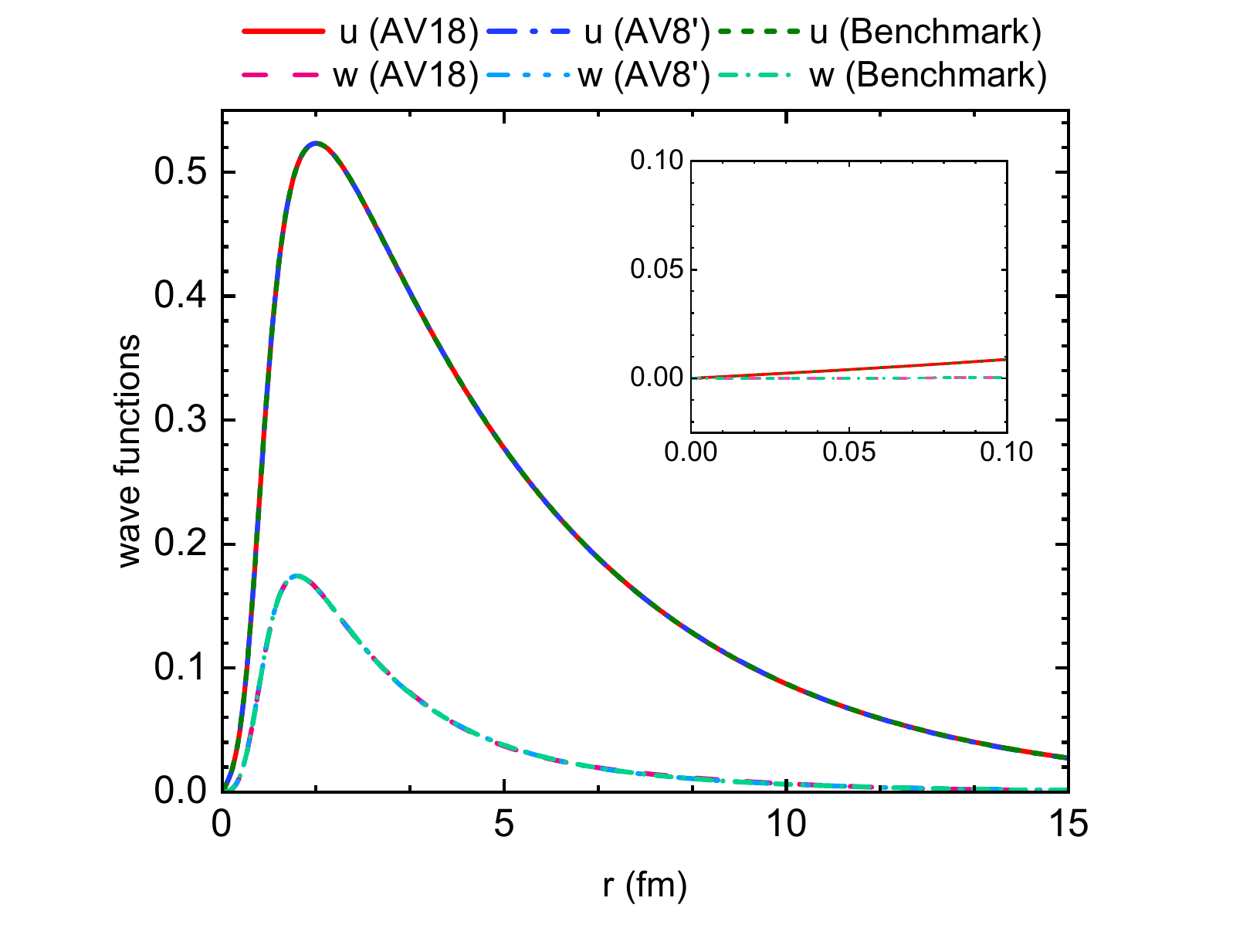}
  \caption{
    Deuteron wave functions with the AV18 and AV8\ensuremath{'} potentials,
    where $ u $ and $ w $ denote the $ S $- and $ D $- wave components, respectively.
    The benchmark is based on the AV18 potential \cite{
      Wiringa1995Phys.Rev.C51_38}
    taken from Ref.~\cite{
      AV18_Web}.
    The hidden layers of each output are composed of three layers each of which has $ 16 $ units.
    Both results the AV18 and AV8\ensuremath{'} potentials show good agreement with the benchmark.}
  \label{wave function}
\end{figure}
\begin{figure}[tb]
  \includegraphics[width=1.0\linewidth]{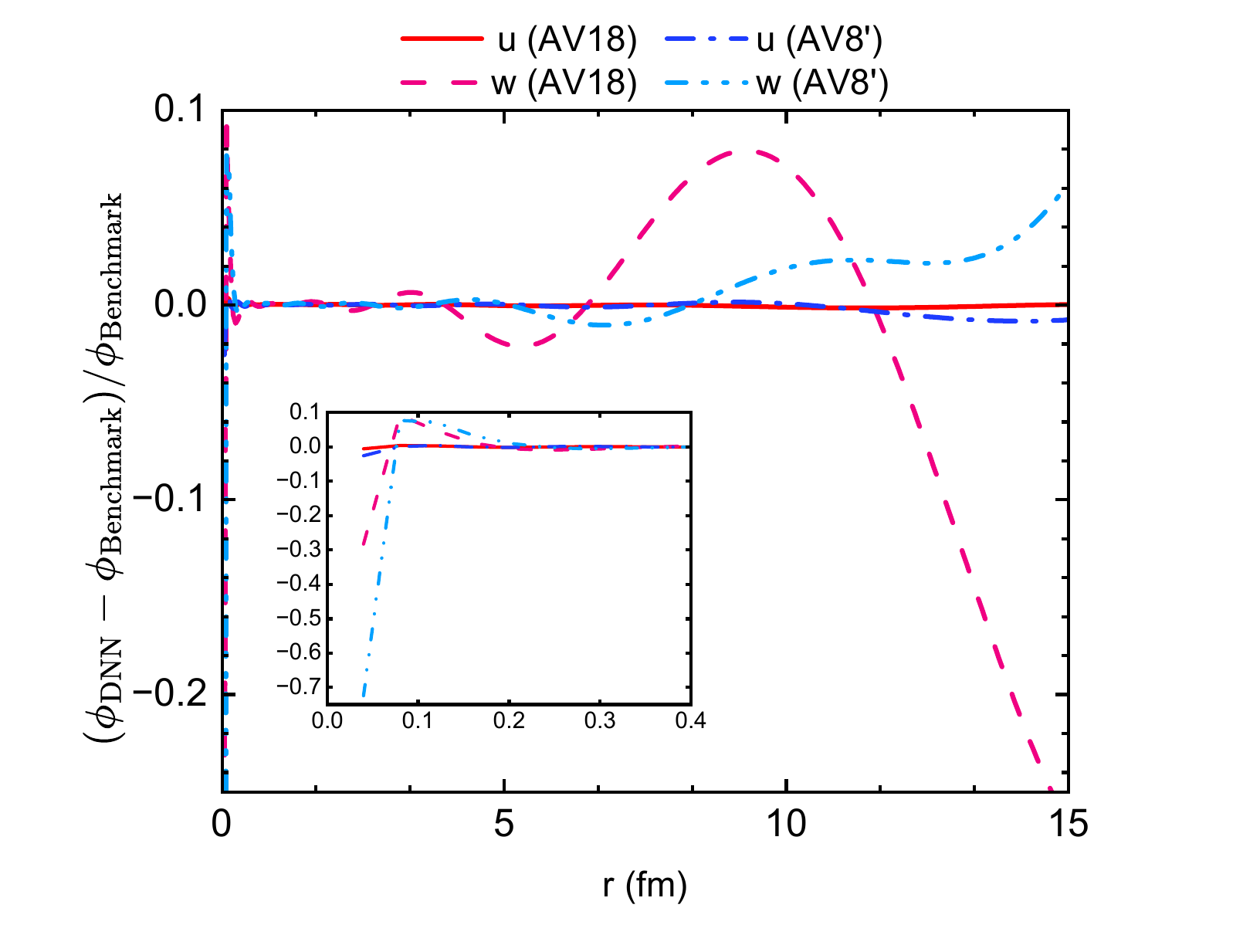}
  \caption{
    Relative error of DNN deuteron wave functions to the AV18 benchmark results \cite{
      Wiringa1995Phys.Rev.C51_38,
      AV18_Web}.
    The hidden layers of each output are composed of three layers each of which has $ 16 $ units.}
  \label{relative wave function}
\end{figure}

Since it is proved that only the $ {}^3 S_1 $ and $ {}^3 D_1 $ states contribute to the ground-state energy,
we can reduce the number of outputs from 18 to 2 to reduce the size of the DNN.
Accordingly, the calculation cost is also reduced;
hence, hereinafter we apply a more complicated hidden layer structure that is composed of three layers each containing $ 16 $ units with $ 30 \, \mathrm{fm} $ box size and 2000 mesh points.
The energies obtained with the AV18, AV8\ensuremath{'}, and AV4\ensuremath{'} potentials are listed in Table~\ref{tab:table2}.
In order to see the effect of the electromagnetic interaction between a proton and a neutron,
calculations using the AV8\ensuremath{'} potential with and without the electromagnetic interaction are also shown.
The accuracy is both within $ 2 \, \mathrm{keV} $ ($ 0.1 \, \% $) for AV18 and AV8\ensuremath{'} calculations compared to the benchmark \cite{
  Wiringa1995Phys.Rev.C51_38}.
AV4\ensuremath{'} potential comprises only four terms; hence, it yields a final error of approximately $ 1 \, \% $ \cite{
  Wiringa2002Phys.Rev.Lett.89_182501}.
In Fig.~\ref{wave function}, the deuteron wave functions are shown based on AV18 and AV8\ensuremath{'} potentials,
which are compared with deuteron wave functions of the AV18 calculation in Ref.~\cite{
  Wiringa1995Phys.Rev.C51_38}.
It is observed that the DNN results of both AV18 and AV8\ensuremath{'} calculations reproduce the AV18 benchmark \cite{
  Wiringa1995Phys.Rev.C51_38} quite nicely.
The relative errors
$ \left[ u^{\urm{DNN}} \left( r \right) - u^{\urm{benchmark}} \left( r \right) \right] / u^{\urm{benchmark}} \left( r \right) $
and 
$ \left[ w^{\urm{DNN}} \left( r \right) - w^{\urm{benchmark}} \left( r \right) \right] / w^{\urm{benchmark}} \left( r \right) $
are shown in Fig.~\ref{relative wave function},
where $ u $ and $ w $ are the partial wave functions of $ {}^3 S_1 $ and $ {}^3 D_1 $ states, respectively.
Because the contribution of the $ {}^3 D_1 $ state in the loss function is much smaller than that of $ {}^3 S_1 $, its relative error is accordingly larger.
\par
The DNN energies with respect to different number of mesh points in the AV18 potential are listed in Table~\ref{tab:table4}.
It is found that 500 mesh points are large enough to produce results less than $ 1 \, \% $ relative error, with the fluctuation of DNN training results becomes an important factor on the precision.
\par
Table~\ref{tab:table3} shows the performance of the program by varying the number of layers and nodes with the AV18 potential.
In the case of a single hidden layer containing $ 16 $ units,
the relative error of energy is $ 2.857 \, \% $.
With increasing the number of units to $ 32 $ or adding a hidden layer, the relative errors do not exceed $ 0.1 \, \% $.
A more complicated network structure does not bring higher precision, since the training results fluctuate slightly every time.
This finding differs from that in Ref.~\cite{sarmiento2024machine}, where adding a second layer induces a sharp decrease in model accuracy, while increasing the number of hidden nodes partially compensates for this loss. 
This shows that our DNN model is more stable and can be effectively extended to model complex physical systems, where expressive power necessitates the use of deeper network architectures.
\par
\begin{figure}[tb]
  \includegraphics[width=1.0\linewidth]{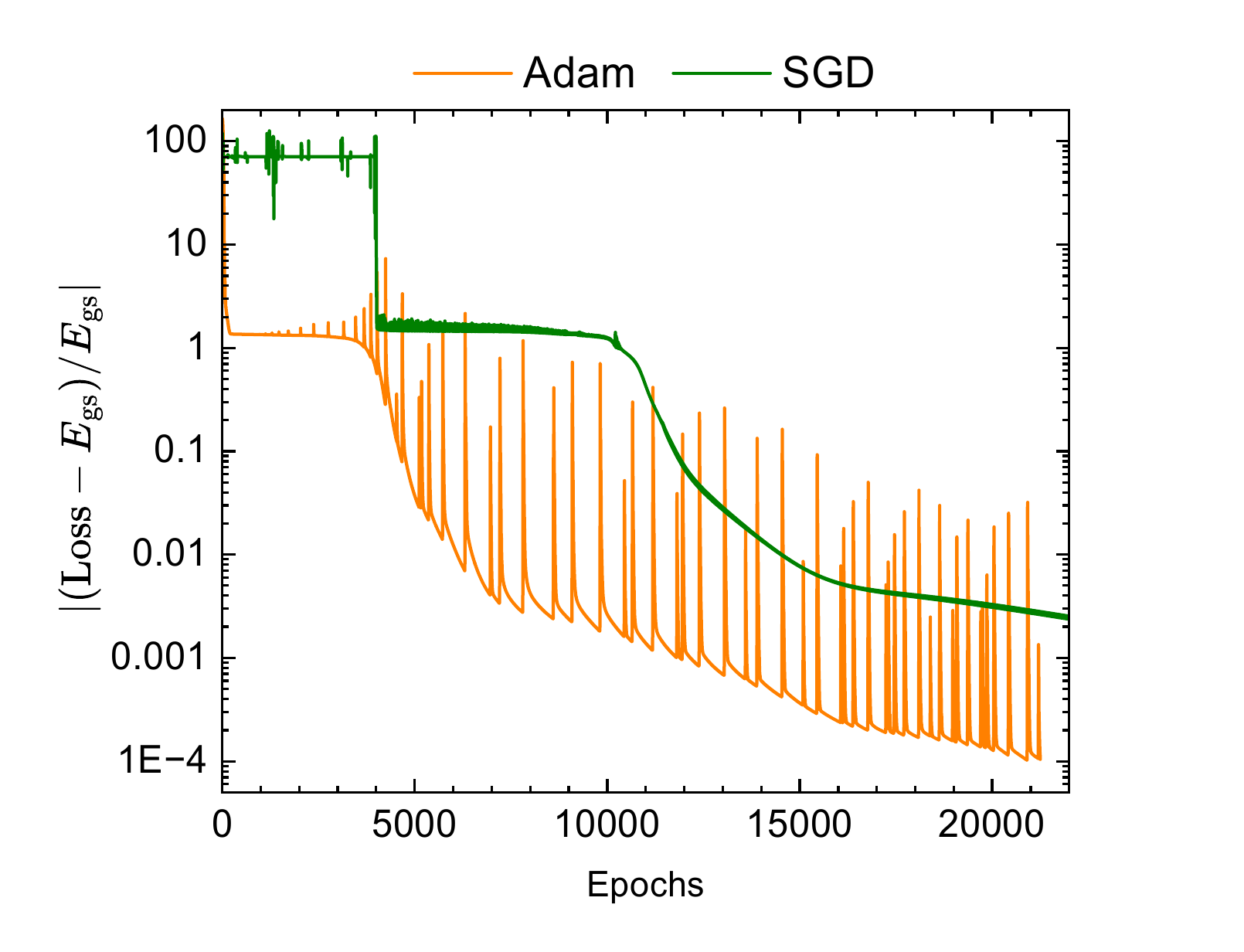}
  \caption{
    Relative error of $ \avr{H} $  with respect to the benchmark AV18 ground-state energy
    $ E_{\urm{gs}} = -2.22458 \, \mathrm{MeV} $~\cite{
      Huang2021Chin.Phys.C45_030002,
      Wang2021Chin.Phys.C45_030003}
    as functions of the number of epochs.
    The hidden layers of each output are composed of three layers each of which has $ 16 $ units.
    The DNN with the Adam optimizer converges around 20000 epochs, while the one with the SGD optimizer fails to converge within $10^{5}$ epochs.}
  \label{loss/epoch}
\end{figure}

In Fig.~\ref{loss/epoch}, the relative errors of the loss function with respect to the benchmark energy are shown as functions of epochs in the AV18 calculation for two optimizers.
If we use the Adam optimizer, spikes appear in the loss values, although the loss still decreases between them.
In contrast, when we use the stochastic gradient descent (SGD) optimizer~\cite{
  10.1214/aoms/1177729586}
shown, these spikes do not appear, but instead it requires more steps to reach the convergence.
\begin{table*}[tb]
  \centering
  \caption{
    The percentage of 18 partial wave states to wave function with the AV18 potential.
    The ground-state energy reads $ -2.2241 \, \mathrm{MeV} $ and $ -2.2246 \, \mathrm{MeV} $ for the non-fully and fully connected neural networks, with both within $ 0.03 \, \% $ relative error to the benchmark value \cite{
      Wiringa1995Phys.Rev.C51_38}.
    Each unconnected hidden layers are two layers,
    each of which is composed of $ 16 $ units.}
  \label{tab:table1}
  \begin{ruledtabular}
    \begin{tabular}{llllll}
      & & \multicolumn{2}{c}{Non-fully connected} & \multicolumn{2}{c}{Fully connected} \\
      & & \multicolumn{1}{c}{$ S = 1 $} & \multicolumn{1}{c}{$ S = 0 $} & \multicolumn{1}{c}{$ S = 1 $} & \multicolumn{1}{c}{$ S = 0 $} \\
      \hline
      $ L = 0 $
      & $ J = 0 $ &                            & $ 7.4428 \times 10^{-9} $  &                            & $5.3286 \times 10^{-9} $  \\
      & $ J = 1 $ & $ 0.9423 $                 &                            & $ 0.9422 $                 &                           \\
      \hline
      & $ J = 0 $ & $ 8.5079 \times 10^{-11} $ &                            & $ 8.3838 \times 10^{-10} $ &                           \\
      $ L = 1 $
      & $ J = 1 $ & $ 9.3388 \times 10^{-8} $  & $ 3.5507 \times 10^{-11} $ & $ 8.8066 \times 10^{-10} $ & $ 9.1213 \times 10^{-9} $ \\
      & $ J = 2 $ & $ 2.6372 \times 10^{-9} $  &                            & $ 2.6894 \times 10^{-7} $  &                           \\
      \hline
      & $ J = 1 $ & $ 0.0577 $                 &                            & $ 0.0578 $                 &                           \\
      $ L = 2 $
      & $ J = 2 $ & $ 4.3943 \times 10^{-11} $ & $ 4.6278 \times 10^{-12} $ & $ 7.2171 \times 10^{-8} $  & $ 1.1914 \times 10^{-8} $ \\
      & $ J = 3 $ & $ 3.2904 \times 10^{-8} $  &                            & $ 9.3754 \times 10^{-9} $  &                           \\
      \hline
      & $ J = 2 $ & $ 1.4599 \times 10^{-9} $  &                            & $ 4.0260 \times 10^{-9} $  &                           \\
      $ L = 3 $
      & $ J = 3 $ & $ 1.0769 \times 10^{-11} $ & $ 4.8327 \times 10^{-9} $  & $ 9.0229 \times 10^{-9} $  & $ 3.1785 \times 10^{-9} $ \\
      & $ J = 4 $ & $ 1.4207 \times 10^{-9} $  &                            & $ 8.0821 \times 10^{-9} $  &                           \\
      \hline
      & $ J = 3 $ & $ 2.3481 \times 10^{-9} $  &                            & $ 4.2018 \times 10^{-9} $  &                           \\
      $ L = 4 $
      & $ J = 4 $ & $ 1.8652 \times 10^{-9} $ & $ 5.2457 \times 10^{-9} $   & $ 4.8368 \times 10^{-9} $  & $ 4.8066 \times 10^{-9} $ \\
      & $ J = 5 $ & $ 6.4078 \times 10^{-9} $ &                             & $ 1.0747 \times 10^{-7} $  &                           \\
    \end{tabular}
  \end{ruledtabular}
\end{table*}
\begin{table}[tb]
  \centering
  \caption{
    DNN results for the energies of the deuteron in the AV18,
    AV8\ensuremath{'} (with and without the electromagnetic (EM) interaction),
    and AV4\ensuremath{'} potentials.
    There exist about $ 1 \, \% $ relative errors in the AV8\ensuremath{'} (without the EM)
    and AV4\ensuremath{'} results \cite{Wiringa2002Phys.Rev.Lett.89_182501}.
    The hidden layers of each output are composed of three layers each of which has $ 16 $ units.
    The benchmark energy is taken from Ref.~\cite{
      Wiringa1995Phys.Rev.C51_38}.}
  \label{tab:table2}
  \begin{ruledtabular}
    \begin{tabular}{lll}
      & \multicolumn{1}{c}{Energy ($ \mathrm{MeV} $)} & \multicolumn{1}{c}{Relative error} \\
      \hline
      Benchmark (AV18)~\cite{Wiringa1995Phys.Rev.C51_38} & $ -2.2246 $ & \\
      \hline
      AV18 (with EM)                 & $ -2.2258 $ & $ 0.054 \, \% $  \\
      AV8\ensuremath{'} (with EM)    & $ -2.2259 $ & $ 0.058 \, \% $  \\
      AV8\ensuremath{'} (without EM) & $ -2.2434 $ & $ 0.845 \, \% $ \\
      AV4\ensuremath{'} (without EM) & $ -2.2436 $ & $ 0.854 \, \% $ \\
    \end{tabular}
  \end{ruledtabular}
\end{table}
\begin{table}[tb]
  \centering
  \caption{
    DNN energies with respect to different numbers of mesh points in the AV18 potential.
    500 mesh points are suffice to generate results within $ 1 \, \% $ relative error, while the relative error barely reduces as the mesh points increasing from $ 1200 $ to $ 1500 $.}
  \label{tab:table4}
  \begin{ruledtabular}
    \begin{tabular}{lll}
      Number of mesh points & Energy ($ \mathrm{MeV} $) & Relative error \\
      \hline
      200  & $ -2.3659 $ & $ 2.729 \, \% $ \\
      500  & $ -2.2248 $ & $ 0.008 \, \% $\\
      1000 & $ -2.2297 $ & $ 0.229 \, \% $\\
      1500 & $ -2.2265 $ & $ 0.085 \, \% $\\
      2000 & $ -2.2258 $ & $ 0.054 \, \% $\\
      \hline
      Benchmark (AV18)~\cite{Wiringa1995Phys.Rev.C51_38} & $ -2.2246 $ & \\
    \end{tabular}
  \end{ruledtabular}
\end{table}
\begin{table}[tb]
  \centering
  \caption{
    Performance test of the deuteron calculation in the AV18 potential.
    Row with ``---'' in the column ``\# of unit in layers'' represents an empty layer.
    The benchmark values with the AV18~\cite{
      Wiringa1995Phys.Rev.C51_38}
    and the experimental data~\cite{
      Huang2021Chin.Phys.C45_030002,
      Wang2021Chin.Phys.C45_030003}
    are listed in the last row.}
  \label{tab:table3}
  \begin{ruledtabular}
    \begin{tabular}{cccccc}
      \multicolumn{3}{c}{\# of unit in layers} & Energy & Relative error & D state prob. \\
      \cline{1-3} \cline{4-4}
      1st & 2nd & 3rd & ($ \mathrm{MeV} $) & &  \\
      \hline
      $ 16 $ & ---    &  ---   & $ -2.16215 $ & $ 2.857 \, \% $ & $ 5.70 \, \% $ \\
      $ 32 $ & ---    &  ---   & $ -2.22552 $ & $ 0.042 \, \% $ & $ 5.77 \, \% $ \\
      $ 16 $ & $ 16 $ &  ---   & $ -2.22536 $ & $ 0.035 \, \% $ & $ 5.77 \, \% $ \\
      $ 16 $ & $ 32 $ &  ---   & $ -2.22588 $ & $ 0.058 \, \% $ & $ 5.77 \, \% $ \\
      $ 32 $ & $ 16 $ &  ---   & $ -2.22569 $ & $ 0.049 \, \% $ & $ 5.77 \, \% $ \\
      $ 32 $ & $ 32 $ &  ---   & $ -2.22580 $ & $ 0.055 \, \% $ & $ 5.77 \, \% $ \\
      $ 16 $ & $ 16 $ & $ 16 $ & $ -2.22581 $ & $ 0.055 \, \% $ & $ 5.77 \, \% $ \\
      $ 32 $ & $ 32 $ & $ 32 $ & $ -2.22575 $ & $ 0.053 \, \% $ & $ 5.77 \, \% $ \\
      \hline
      \multicolumn{3}{l}{Expt.~\cite{Huang2021Chin.Phys.C45_030002, Wang2021Chin.Phys.C45_030003}} & $ -2.22458 $  &             &                \\
      \multicolumn{3}{l}{Benchmark (AV18)~\cite{Wiringa1995Phys.Rev.C51_38}}                       & $ -2.22458 $  &             & $ 5.76 \, \% $ \\
    \end{tabular}
  \end{ruledtabular}
\end{table}
% 
%%%%%%%%%%%%%%%%%%%%%%%%%%%%%%%%%%%%%%%%%% 
\section{Summary}
\label{Sect:V}
\par
In Ref.~\cite{
  Naito2023Phys.Rev.Research5_033189},
an unsupervised machine learning technique is developed to calculate the ground state with deep neural network.
In this paper, we extended the method by introducing the spin and isospin degrees of freedom through the introduction of partial wave expansions, which are generated by a non-fully-connected deep neural network.
\par
The method is verified by calculating the simplest two-body nuclear system---deuteron.
At first, the partial waves with the orbital angular momentum quantum number $ L $ from zero to four
and the possible total spin $ S $ from zero to one are calculated.
The results were consistence with the known fact, i.e., only the $ {}^3 S_1 $ and $ {}^3 D_1 $ states contribute to the deuteron ground state.
In the following calculations, the obtained wave functions and energies show consistency with the benchmark \cite{
  Wiringa1995Phys.Rev.C51_38,
  Wiringa2002Phys.Rev.Lett.89_182501}.
We find that the deep neural network does not need to be large.
In the present case, two hidden layers and each of them containing eight units are sufficient to generate faithful representations of the wave function. 
\par
We believe that these improvements to the deep neural network approach hold promise for more studies that could further extend the methodology of this work.
For example, one is to extend from two-body calculations to $ N $-body calculations. 
For three-body systems, we believe the ground state can be obtained by solving the Faddeev equation using hyperspherical coordinates \cite{nielsen2001three}
, in which the corresponding partial wave expansion includes two angular terms with spin and isospin components.
Based on the present work and our deep neural network approach to solve the Dirac equation \cite{wang2025deep}, another promising direction is to incorporate spin and isospin degrees of freedom into the Dirac framework. 
We anticipate that the present deep neural network architecture can be extended to generate Dirac partial-wave functions.
% 
%%%%%%%%%%%%%%%%%%%%%%%%%%%%%%%%%%%%%%%%%%%%%%%%%% 
% 
\begin{acknowledgments}
  The authors thank Koji Hashimoto and Hisashi Naito for the fruitful discussion.
  C.W.~acknowledges the warm hospitality of the RIKEN iTHEMS Center.
  T.N.~acknowledges
  the JSPS Grant-in-Aid for Transformative Research Areas (A) under Grant No.~JP25H01558,
  the JSPS Grant-in-Aid for Scientific Research (S) under Grant No.~JP25H00402,
  the JSPS Grant-in-Aid for Scientific Research (B) under Grant Nos.~JP23K26538 and JP25K01003,
  the JSPS Grant-in-Aid for Scientific Research (C) under Grant No.~JP23K03426,
  the JSPS Grant-in-Aid for Early-Career Scientists under Grant No.~JP24K17057,
  the JSPS Grant-in-Aid for JSPS Fellows under Grant No.~JP25KJ0405,
  and
  JST COI-NEXT Grant No.~JPMJPF2221.
  J.L.~acknowledges
  the National Natural Science Foundation of China (Nos.~12475119 and 11675063) and
  the Key Laboratory of Nuclear Data Foundation (JCKY2025201C154).
  H.L.~acknowledges the JSPS Grant-in-Aid for Scientific Research (S) under Grant No.~JP20H05648 and the RIKEN Pioneering Project: Evolution of Matter in the Universe.
  The numerical calculations were performed on cluster computers at the RIKEN iTHEMS Center.
\end{acknowledgments}
%%%%%%%%%%%%%%%%%%%%%%%%%%%%%%%%%%%%%%%%%%%%%%%%%%%%%%%%%%% 
% 
\appendix
\section{DNN output analysis}
\label{Appen:A}
\par
\begin{figure}[tb]
  \includegraphics[width=1.0\linewidth]{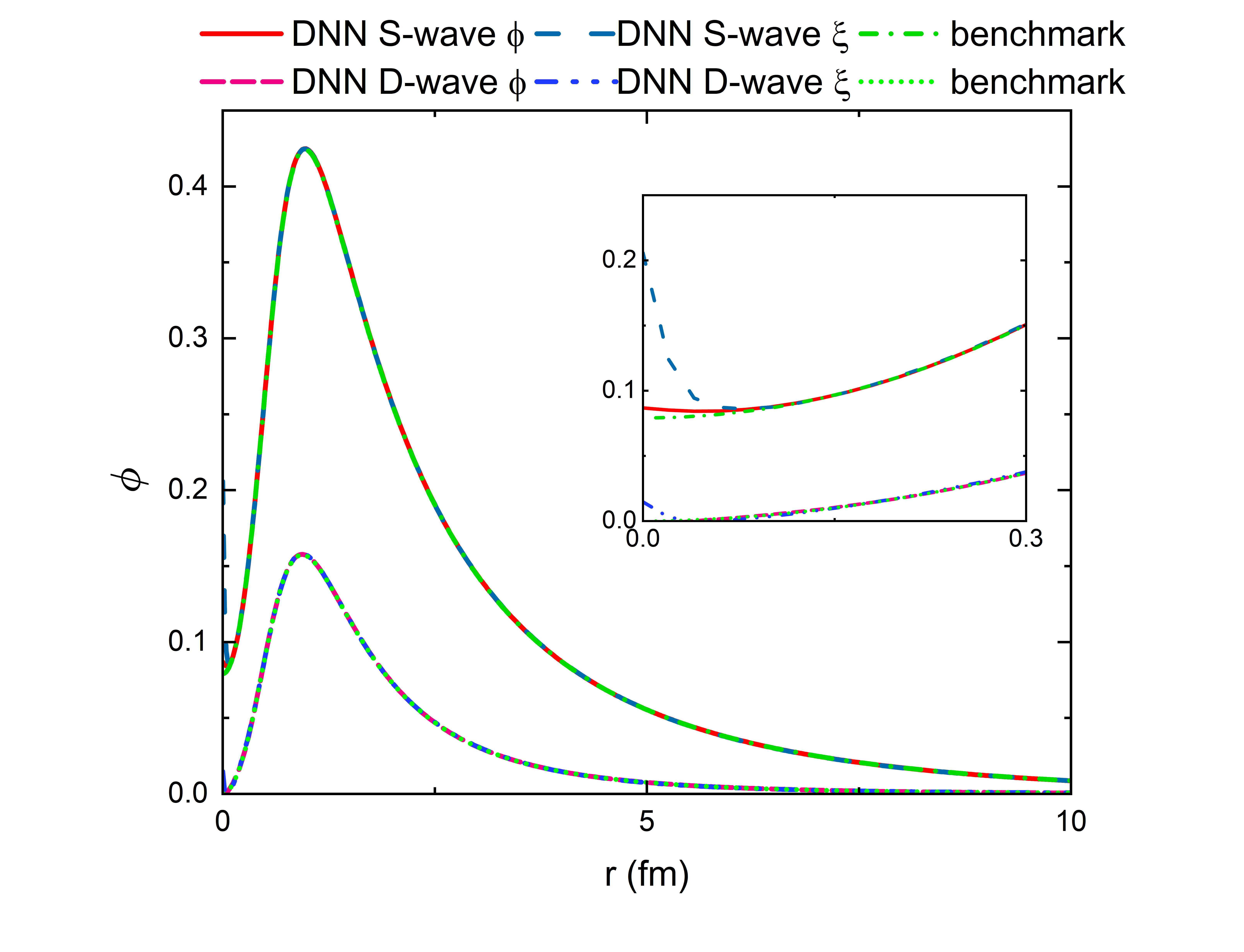}
  \caption{
    Wave functions $ \varphi_{LSJ} $ obtained by
    1) the output corresponds to $ \xi_{LSJ} $ (Blue) and
    2) the output corresponds to $ \varphi_{LSJ} $ and $ \xi_{LSJ} $ is calculated from the output (Red).
    The benchmark calculation~\cite{Wiringa1995Phys.Rev.C51_38} is also shown (Green).}
  \label{wave function origin}
\end{figure}

We show the wave functions corresponding to two strategies of selecting DNN outputs: 
1) the output corresponds to $ \xi_{LSJ} $ and
2) the output corresponds to $ \varphi_{LSJ} $ and $ \xi_{LSJ} $ is calculated from the output as did in the main text.
In our test, we use $ 1700 $ mesh points in a $ 20 \, \mathrm{fm} $ box to generate $ {}^3 S_1 $ and $ {}^3 D_1 $ states.
We find that the DNN can generate more precise results of $ \varphi_{LSJ} $ than $ \xi_{LSJ} $ close to the origin as shown in Fig.~\ref{wave function origin}.
Therefore, $ \varphi_{LSJ} $ is utilized as the DNN output in our calculation.
\end{document}